\documentclass[prl,aps,tightenlines,twocolumn,nofootinbib]{revtex4}
\usepackage{graphicx}

\begin{document}
\newcommand{\be}{\begin{equation}}
\newcommand{\ee}{\end{equation}}
\newcommand{\mx}{\mbox}
\newcommand{\mt}{\mathtt}
\newcommand{\mtf}{\mathtt{f}}
\newcommand{\mts}{\mathtt{s}}
\newcommand{\p}{\partial}
\newcommand{\st}{\stackrel}
\newcommand{\al}{\alpha}
\newcommand{\bb}{\beta}
\newcommand{\ga}{\gamma}
\newcommand{\te}{\theta}
\newcommand{\de}{\delta}
\newcommand{\et}{\eta}
\newcommand{\ze}{\zeta}
\newcommand{\s}{\sigma}
\newcommand{\e}{\epsilon}
\newcommand{\om}{\omega}
\newcommand{\Om}{\Omega}
\newcommand{\la}{\lambda}
\newcommand{\La}{\Lambda}
\newcommand{\ti}{\widetilde}
\newcommand{\ih}{\widehat{i}}
\newcommand{\jh}{\widehat{j}}
\newcommand{\kh}{\widehat{k}}
\newcommand{\lh}{\widehat{l}}
\newcommand{\eh}{\widehat{e}}
\newcommand{\ph}{\widehat{p}}
\newcommand{\qh}{\widehat{q}}
\newcommand{\mh}{\widehat{m}}
\newcommand{\nh}{\widehat{n}}
\newcommand{\Dh}{\widehat{D}}
\newcommand{\2}{\frac{1}{2}}
\newcommand{\3}{\textstyle{1\over 3}}
\newcommand{\4}{\frac{1}{4}}
\newcommand{\8}{\textstyle{1\over 8}}
\newcommand{\6}{\textstyle{1\over 16}}
\newcommand{\ra}{\rightarrow}
\newcommand{\lra}{\longrightarrow}
\newcommand{\Ra}{\Rightarrow}
\newcommand{\im}{\Longleftrightarrow}
\newcommand{\hs}{\hspace{5mm}}
\newcommand{\vs}{\vspace{5mm}\\}

\title{Can we have a stringy origin behind $\Omega_{\Lambda}(t)\propto
\Omega_{m}(t)$?}  
\author{Tirthabir Biswas and Anupam Mazumdar}
\affiliation{CHEP, McGill University, Montr\'eal, QC, H3A~2T8,
Canada.}

\begin{abstract}
Inspired by the current observations that the ratio of the abundance
of dark energy $\Omega_{\Lambda}$, and the matter density,
$\Omega_{m}$, is such that $\Omega_{m}/\Omega_{\Lambda}\sim 0.37$, we
provide a string inspired phenomenological model where we explain this
order one ratio, the smallness of the cosmological constant, and also
the recent cosmic acceleration. We observe that any effective theory
motivated by a higher dimensional physics provides radion/dilaton
couplings to the standard model and the dark matter component with
different strengths.  Provided radion/dilaton is a dynamical field we
show that $\Omega_{\Lambda}(t)$ tracks $\Omega_{m}(t)$ and dominates
very recently.
\end{abstract}

\maketitle

The cosmological constant problem is one of the most difficult
problems of theoretical physics. What requires an explanation is the
fact that the observed energy content of the Universe, $\sim 4\times
10^{-47}~({\rm GeV})^4$, is many orders of magnitude smaller than that
of the theoretical prediction for the cosmological constant energy
density alone. The mismatch is of order $10^{-120}\times M_{p}^4$,
where $M_{p}\sim 2.4\times 10^{18}$~GeV.  One would naively expect
that even if the bare cosmological constant can be made to vanish, the
quantum corrections would eventually lead to quadratic divergences
$\sim M_{p}^4$. This is known as why the bare cosmological constant is
so small and stable under quantum corrections~\cite{Weinberg}.  One
can resort to supersymmetry which makes it possible to keep the vacuum
energy density under control, but which, nevertheless, has to be
broken at weak scale. Therefore invoking supersymmetry certainly
ameliorates the problem by $10^{68}$ orders of magnitude, but does not
solve it at all.

There is also a kind of a coincidence problem, sometimes dubbed as a
why now problem. Recent observations from supernovae~\cite{SN} and
from the cosmic microwave background (CMB) anisotropy
measurements~\cite{WMAP} suggest that the majority of the energy
density $\sim 70 \%$ is in the form of dark energy, whose constituent
is largely unknown, but usually believed to be the {\it cosmological
constant} with an equation of state $\omega=-1$, which is also
responsible for the current acceleration. In this respect not only the
cosmological constant is small, but it also happens to be dominating
the energy density of the Universe.  In principle the physics behind
the value of the cosmological constant and its ``evolution'' need not
be related to the redshifting of matter/radiation density. Is it then
just by coincidence that today $\rho_{\La}$ is close to the value
$\rho_m$ or are there deeper connections between the two? A related
question would be why the cosmological constant is dominating right
now and for example not during the big bang nucleosynthesis (BBN) at a
temperature $T\sim 1$~MeV?

Both the smallness and the why now questions can be answered in part
if we believe that the physics of the dark energy is somehow related
to the rest of the energy density of our Universe. In some sense the
dark energy needs to track the infrared part of gravity which is
responsible for the expansion of the Universe. Attempts have been made
to construct such tracking mechanisms in single and multi fields, such
as dynamical quintessence \cite{Steinhardt}, k-essence~\cite{Mukhanov}
type models, non-Abelian vacuum structure with non-vanishing winding
modes~\cite{Jaikumar}, modified Friedmann equation at late
time~\cite{Freese}, or large distances~\cite{Dvali}, rolling dilaton
with dilatonic dark matter~\cite{Amendola}, or due to inflationary
backreaction~\cite{Brandenberger}. Only in the latter reference it was
argued that $\Omega_{\Lambda_{eff}}(t)\sim {\cal O}(1)$ throughout the
evolution.

In this paper we propose a simple model which arises naturally from
the compactification of a higher dimensional theory. In string theory
compactification it is known that the dilaton and various moduli
fields like radion, shape moduli couple to the Standard Model (SM)
degrees of freedom. For simplicity we may assume that all but a single
combination, $Q$, of various moduli has been stabilized at higher
scales, see for instance \cite{t}. In an effective four dimensional
theory $Q$ field couples to the SM degrees of freedom and also to
the dark matter. We assume that at least the SM and some relevant
(beyond SM) degrees of freedom have already been thermally/non-thermally
excited in the early Universe, e.g.~see Ref.~\cite{Enqvist}. In
general the potential for $Q$ has a run away behavior,
see~\cite{Witten}, nevertheless we will show that, by virtue of its
coupling to the matter field ``the effective potential'' for $Q$ can
have a local minimum which always tracks the matter density of the
Universe. Further observe that the $Q$ field in general will have
different couplings to fermions, scalars and gauge bosons by virtue 
of its running below the string compactification scale~\cite{Polyakov}.

Let us imagine that our world were originally higher dimensional, such
that the three spatial dimensions, along with the origin of the SM are
all due to some interesting compactification of the extra spatial
dimensions. In the Einstein frame the most generic action will take a
form
$S=S_{\mt{grav}}+S_{\mt{fermions}}+S_{\mt{gauge~bosons}}+S_{\mt{scalars}}$.
where
\be 
S_{\mt{grav}}=M_p^2\int d^4x\ \sqrt{-g}\left[R/2- (\p Q)^{2}/2 
- V(Q)\right]\,, 
\ee
\be 
\label{ferm}
S_{\mt{fermions}}=\int d^4x\
\sqrt{-g}\bar{\psi_{\mtf}}\left(\nabla{\hskip-3.0mm}/+im_{\mtf}e^{2\mu_{\mtf}
Q}\right)\psi_{\mtf}\,, 
\ee
where the coupling exponent $\mu_{\mtf}$ are different for
different species, see Ref.~\cite{Polyakov}.  Similarly the most
generic coupling of $Q$ to the gauge bosons can be written as
\be
\label{gbos}
S_{\mt{gauge~bosons}}=({1}/{4\pi\al_0})\int d^4x \sqrt{-g}e^{2\mu_r Q}F^2\,.
\ee 
From (\ref{gbos}) we can read the fine structure constant to be $\sim
\al_0e^{-2\mu_r Q}$. Finally we discuss the fundamental scalar
coupling:
\be 
\label{scal}
S_{\mt{scalars}}=\int d^4x\
\sqrt{-g}\left(-(\p \phi_{\mts})^2/2-m_{\mts}^2
e^{4\mu_{\mts} Q}\phi_{\mts}^2\right)\,. 
\ee 
The radiation content of the Universe is determined by the SM
relativistic degrees of freedom, while the cold dark matter component
is arising from either scalars, fermions or gauge bosons beyond the SM
gauge group, and the dark energy density is identified by,
\be \ti{\rho}_{d}=\dot{Q}^2/2+V_{\mt{eff}}(Q)\,, 
\ee
where the effective dark energy potential is given by,
\be
\label{eff} 
V_{\mt{eff}}(Q)=V(Q)+\sum_{i=\mts,\mtf,r}e^{2\mu_i Q}\rho_i\,, 
\ee 
with an individual component of matter and radiation density
determined by $\rho_{i}$. Given all the components we can write down
the Friedmann equation for a Robertson Walker metric,
\begin{eqnarray}
\label{Hubble}
H^2 &=& ({1}/{3M_{p}^2})\ti{\rho}_t=({1}/{3M_{p}^2})
\sum_{I=\mts,\mtf,r,d}\ti \rho_{I}\,, \nonumber \\
&=&({1}/{3})\left(\sum_i e^{2 \mu_i Q}\rho_i+\dot{Q}^2/2+V(Q)\right)\,.
\end{eqnarray} 
The evolution equations for $Q,~\rho_{i}$ read as,
\begin{eqnarray}
\label{eqm}
\ddot{Q}+3H\dot{Q}=-(V'(Q)+2\sum_i\mu_ie^{2\mu_i Q}\rho_i)\,, \\
\label{eqm1} 
\dot{\rho_i}+3H(p_i+\rho_i)=0 \,.
\end{eqnarray} 
for an equation of state, $p_i=\om_i \rho_i$. Note that although
$\ti{\rho}_i$ is the energy density we measure, it is the bare energy
density, $\rho_i$, which obeys the usual equations of state. This can
be seen for example in the case of a non-relativistic dust. The
density is given by $e^{2\mu_i Q}m_iN/V\equiv e^{2\mu_i Q}\rho_i$,
where $N$ is the total number of particles, a constant, and $V$ is the
volume of the universe, which redshifts as $a^{-3}$ leading to
$\rho_i\sim a^{-3}$. From Eq.~(\ref{eqm1}) we obtain the standard result
$\rho_i=\rho_{0i}\left({a}/{a_0}\right)^{-3(1+\om_i)}$.

Let us now assume a generic potential for the field $Q$.  Keeping in
mind that $Q$ could either be a dilaton/radion or some linear
combination of various moduli, such as a volume and/or shape
moduli, the potential can be written as
\be
\label{potQ}
V(Q)=V_0e^{-2\bb Q}\,, 
\ee 
where we assume $V_{0}$ is the scale of new physics, which could be
determined by the string or the Planck scale.  Such a run away
potential for $Q$ may arise very naturally for the radion/dilaton
field from the fluxes, or the p-brane contribution, or from the
internal curvature of the compactified manifolds.

Let us also assume that there are two dominating components in the
Universe, the dark energy determined by $Q$ and the matter
component. Further note, if $\mu$'s and $\bb$ have the same sign (we
will furnish examples later where this can happen) then the effective
potential for the $Q$ field, see Eqs.~(\ref{eff},\ref{potQ}), provides
a dynamical stabilizing mechanism, because the two overall exponents
differ in signs. In this case the motion of $Q$ field will always
track the minimum of the potential Eq.~(\ref{eff}), and therefore if
the evolution of the Universe is adiabatic, the motion of $Q$ will be
as well, provided $\beta\geq {\cal O}(1)$.

Let us describe the evolution of $Q$ field in an adiabatic
approximation, where we neglect the higher time derivatives of $Q$,
e.g. $\ddot Q \ll 3H\dot Q$. This assures that the $Q$ field always
stays at the minimum of the ``effective potential'' for which the
evolution of $Q$ can be obtained in terms $\rho$. $V_{\mt{eff}}'(Q)=0$
implies
\be 
 e^{2Q}=\left({\bb V_0}/{\mu_{i}\rho_{i}}\right)^{1/(\mu_{i}+\beta)}
\,.  
\ee 
After some algebra we can evaluate the fraction of the
radiation/matter density with respect to the critical 
density of the Universe
\be
\label{omega}
\Omega_{i}={\ti{\rho}_{i}}/{\ti{\rho}_t} = \bb/({\mu_{i}+\bb})\,,
\ee
where $i$ corresponds to either radiation or dark matter depending on
the epochs we are interested in.  Within an adiabatic approximation we
can also estimate the Hubble expansion rate,
\begin{eqnarray}
\label{Hub}
H^2 &=&V_0\left(1+{\bb}/{\mu_{i}}\right)\left(
{\bb V_0}/{\mu_{i}}\right)^{-\bb/(\mu_{i}+\bb)}
\rho^{\bb/(\mu_{i}+\bb)}\,, \\
H&\sim& a^{-{(3/2)(1+\om_i)}\bb/(\mu_{i}+\bb)}\,.
\end{eqnarray}
The above equations can be solved to give
\be
\label{scalefact}
a(t)=a_0\left({t}/{t_0}\right)^{(2/3\beta)
\left[(\mu_{i}+\bb)/(1+\om_i)\right]}\,.
\ee
Note that the scale factor depends on the ratio, $\mu_{i}/\beta$,
and if $\mu_{i}/\beta < 1$ then we follow a natural course of
radiation or matter dominated epochs.

If we are in a matter dominated epoch, such that $\omega_c=0$, where
$c$ stands for the cold dark matter component, then we obtain an
interesting relationship in order to have an accelerated expansion
\begin{equation}
\label{res1}
{\mu_{c}}/{\beta}> {1}/{2}\,.
\end{equation}
Now if we take the matter component, $\Omega_{c}\sim 0.27$, we obtain
from Eq.~(\ref{omega}), $\mu_{c}/\beta\sim 2.7$, indicating an
accelerated expansion. In other words $\mu_{c}/\beta\sim 2.7$ explains
the value of $\ti{\rho}_{d}\sim 2.7\ti{\rho}_c\sim 10^{-120}M_p^4$,
or, $\Omega_{c}/\Omega_{d}\sim 0.37$. Further note that the equation
of state for the dark energy comes out to be, $\omega_Q\sim -1$,
supports the current observation,~\cite{SN}, by virtue of $\dot Q^2/2
\ll V(Q)$.

Similar calculation can be repeated during the radiation epoch,
$(\omega_r=1/3)$, by assuming that the dominant component of the energy
density is in the form of radiation and the coupling of the $Q$ field
to radiation is determined by the exponent $\mu_{r}$ (for simplicity
we can assume that for all the relativistic degrees of
freedom, $\mu_{i} =\mu_{r}$, though this is not necessary), then we
obtain
\begin{equation}
\label{res2}
{\mu_{r}}/{\beta}> 1\,,
\end{equation}
in order to have an accelerated expansion. However during BBN the
Hubble parameter is well constrained.  One could allow the dark energy
component, $\Omega_{d} < 0.13$ \cite{Joyce}, while $\Omega_{r}> 0.87$,
which provides the ratio $\mu_{r}/\beta \sim 0.14$.  Indeed the ratio
turns out to be smaller than one, indicating a decelerating
expansion. However as we shall see below the value of $\mu_{r}$ has to
be even smaller.

From the above analysis it may appear that the moment we enter the
matter dominated era, the Universe starts to accelerate, but this is
not correct. First of all our Eqs.~(\ref{omega},\ref{scalefact})
should be only viewed as asymptotic expressions when the two component
approximation is valid. However during the transition from radiation
to cold dark matter, the evolution in general will be complicated. For
example, let us assume, $\mu_c\gg\mu_r$, note that in our case $\ti
\rho_{b},\ti\rho_{c}$ redshift differently, indicating that baryons,
$\ti\rho_{b}$ were dominating over the CDM not so long ago. In fact we
can compute approximately when the CDM began to dominate over the
baryons, which will also roughly correspond to the beginning of an
accelerated expansion,
\be
1+z_{\mt{accel}}=\left({\Om_{c_0}}/{\Om_{b_0}}\right)^
{\left[(\mu_c+\bb)/3\mu_c\right]}\,.
\ee 
Taking $\mu_{c}/\beta\sim 2.7$, as obtained above, we find
$z_{\mt{accel}}\approx 1.8$. Further note that the above interesting
result is determined solely by the ratios of
$\Om_{d_0}/\Om_{c_0}$ and $\Om_{b_0}/\Om_{c_0}$.

However in this setup we end up with more baryons than CDM at early
epochs, say during the CMB formation. In order to address this issue,
we will have to assume here two competing dark matter candidates, one
with $\ti\rho_{w}$, with a different coupling to $Q$, say $\mu_{w}$
and the other $\ti\rho_{c}$ with $\mu_{c}$.  For the purpose of
illustration we may as well assume; $\mu_{w}= \mu_{r}\ll \mu_{c}$, and
at present $\ti\rho_{w}\sim \ti\rho_{c}$. However note that for
$\mu_{w}=\mu_{r}$, $\ti\rho_{w}$ redshifts according to the baryons,
therefore maintaining a constant ratio throughout the evolution. On
the other hand, $\ti\rho_{c}$, by virtue of its coupling to $Q$,
becomes $\ti\rho_{c} \ll \ti\rho_{w}$, at early times. In this
scenario we can achieve two things; first maintain an almost constant
baryon to CDM ratio since BBN, and second we obtain the very late
acceleration, e.g. $z_{\mt{accel}}\sim {\cal O}(1)$. Using
$\Omega_{m}\sim 0.27$, and the present baryonic abundance to be
$\Omega_{b}\sim 0.014$, we find that in our case $\ti\rho_{c}\geq
\ti\rho_{w}+\ti\rho_{b}$, at $z_{\mt{accel}}\sim 0.5$. This is the
time when the dark energy starts dominating, and also the accelerating
phase begins.  The baryon abundance at early times, say during CMB is
given by $\Omega_{b}\sim 0.044$, while the ratio of baryonic 
and CDM abundance nearly remains, $\Omega_{b}/\Omega_{m}\sim 0.2$, until
very recently when it decreases slightly as $\ti\rho_{c}\geq \sim \rho_{w}$.

By taking slightly smaller baryonic abundance today, $\Omega_{b}\sim
0.01$, we obtain $z_{\mt{accel}}\sim 1$. These numbers are encouraging
because they are fairly close to the best fit within the present
uncertainties~\cite{WMAP}.

What is crucial for our mechanism to work is the fact that the
coupling exponents, $\mu$'s, have to have signs opposite to that of
$\bb$. Note that one can always redefine $Q$ so that individual signs
do not matter. 

Let us now consider two examples of the mechanism where we identify
$Q$ with the internal volume of the extra dimensional manifold, with
$\bb$ positive in one case and negative in the other. In the first
example, let us assume Eq.(\ref{potQ}) arising from the string
gas~\cite{Vafa}, where the strings are winding one compactified extra
dimension which has a simple topology of a circle. We assume that the
dilaton and the rest of the modulii of the additional extra five
dimensions are assumed to have been stabilized at the string scale via
some mechanism. In this case the string {\it winding modes} of a gas
of strings would exert an effective negative pressure, which would
result in an effective exponential potential for $Q$, where
$\bb<0$~\cite{watson}. On the other hand we may assume that all the
fermions, gauge bosons and the scalars are living in the entire
bulk. Upon dimensional reduction this would lead to negative coupling
exponents in Eqs.~(\ref{ferm},\ref{gbos},\ref{scal}), for example see
Ref.~\cite{anupam}.  In this respect we also assume that the cold dark
matter candidate resides in the bulk, it could be some stable bulk
fermions for instance.

The second example consists of an usual M-theory/Supergravity type
reduction where the potential for $Q$ is either coming from internal
curvature terms or fluxes (see for example \cite{pope,t}), where $\bb$
comes with a positive sign. Now consider the gauge fields to originate
from the off-diagonal components of the metric as in the usual
Kaluza-Klein scenario; then the radiation coupling comes with always
$\mu_r>0$, with an overall positive exponent, see~\cite{kaluza}. In
this respect one could as well imagine the cold dark matter component
originates from the massive gauge bosons of the hidden degrees of
freedom. Such candidates could arise in the mirror Universe,
see~\cite{Zurab}.

Let us now focus on some of the important observational
constraints. Due to the low scales involved, although the background
$Q$ field is stabilized, the $Q$ quanta are virtually massless
($m_Q\sim 10^{-33}$ ev), and hence can mediate a fifth force violating
equivalence principle. Tests on violation of equivalence principle
essentially constraints $\mu<10^{-4}$~\cite{review}, for ordinary
matter and radiation.  Note that the bounds really constrain $\mu$'s
for baryonic matter. Couplings for gauge bosons, leptons and neutrinos
only follow if one assumes the universal coupling as in the case of
Brans-Dicke theory.

In our case we require $\mu_{c}>\beta/2 > \mu_{r}$ and $\beta\geq
{\cal O}(1)$, therefore even if $\mu_r\sim 10^{-4}$, the above
relationship can be satisfied. The running of the coupling constants
due to stringy loop corrections may naturally lead to small values for
$\mu$'s~\cite{Polyakov}. On the other hand, recently it has been also
realized that even if $\mu_r$'s are significantly larger ($\sim 0.1$),
it might avoid any detection in the fifth force experiments due to
``chameleon'' type mechanisms, see~\cite{justin}.

It is evident that in our model the physical constants like fine
structure constant or the gravitational constant (or equivalently the
masses) varies with time through its $Q$-dependence. Using straight
forward algebra we obtain the ratio of the fine structure constants,
\be
{\al_{\mt{bbn}}}/{\al_{\mt{0}}}=
e^{-2\mu_r(Q_{\mt{bbn}}-Q_{\mt{0}})}=
\left({\mu_r\ti{\rho}_{\mt{bbn}}}/
{\mu_c\ti{\rho}_{\mt{0}}}\right)^{\mu_r/\bb}\,,
\ee
where the subscript $0$ corresponds to the present time.  The
variation of $\al$ is constrained to be within a few percent since
BBN, $-0.06<\Delta \alpha/\alpha<0.02$~at $95~\%$ CL~\cite{Martins},
we obtain $\mu_{r}/\beta\sim 10^{-3}$.  Note that the CDM coupling to
$Q$ is unconstrained from the above mentioned experiments. 

Similarly the analysis for variation of $G_{N}$ is almost identical to
$\al$, except that one has to replace $\mu_r$ with $\mu$'s
corresponding to the fermions, see~\cite{review}. Note that
the success of our model relies on, $\mu_{r}\ll\mu_{c}$.

To summarize, we associate the dark energy component of the Universe
by the dynamics of a stringy component which could be a linear
combination of radion, dilaton or shape moduli. We point out that if
such a field couples to the SM relativistic particles differently than
that of the cold dark matter, then it is possible to explain the
recent acceleration during the matter dominated era, close to
$z_{\mt{accel}}\sim {\cal O}(1)$. We also show that the ratio of the
baryons and the CDM abundance can be maintained within the close
proximity of the current observations.

Nevertheless, many interesting questions remain, such as the
fluctuations of $Q$ during its slow roll evolution, the two types of
CDM and their role in galaxy formation. It is also important to go
beyond the two component approximation in the evolution and include
all the matter components.  These questions we leave for future
investigation.

This work is supported in part by the NSERC.  We would like to thank
Luca Amendola and Andrew Liddle for discussion. We are grateful to Guy
Moore for pointing out important caveats in our first version of the
draft.


\end{document}